\documentclass{article}
\usepackage{graphicx}

\begin{document}

\title{Resonant nonlinearity management for nonlinear-Schr\"{o}dinger
solitons}
\author{Hidetsugu Sakaguchi$^{1}$ and Boris A. Malomed$^{2}$ \\
$^{1}$Department of Applied Science for Electronics and Materials,\\
Interdisciplinary Graduate School of Engineering Sciences,\\
Kyushu University, Kasuga, Fukuoka 816-8580, Japan\\
$^{2}$Department of Interdisciplinary Studies,\\
School of Electrical Engineering, Faculty of Engineering,\\
Tel Aviv University, Tel Aviv 69978, Israel}
\maketitle

\begin{abstract}
We consider effects of a periodic modulation of the nonlinearity coefficient
on fundamental and higher-order solitons in the one-dimensional NLS
equation, which is an issue of direct interest to Bose-Einstein condensates
in the context of the Feshbach-resonance control, and fiber-optic
telecommunications as concerns periodic compensation of the nonlinearity.
We find from simulations, and explain by means of a straightforward
analysis, that the response of a fundamental soliton to the weak perturbation 
is resonant, if the modulation frequency $\omega $ is close to the intrinsic frequency of the soliton. For higher-order $n$-solitons with $n=2$ and $3$, the response to
an extremely weak perturbation is also resonant, if $\omega $ is close to
the corresponding intrinsic frequency. More importantly, a slightly stronger
drive splits the $2$- or $3$-soliton, respectively, into a set of two or
three moving fundamental solitons. The dependence of the threshold
perturbation amplitude, necessary for the splitting, on $\omega $ has a
resonant character too. Amplitudes and velocities of the emerging
fundamental solitons are accurately predicted, using exact and approximate
conservation laws of the perturbed NLS equation.
\end{abstract}


\section{Introduction}

The nonlinear Schr\"{o}dinger (NLS) equation is a universal model
of weakly nonlinear dispersive media \cite{Whitham,Zakharov}. The
existence and stability of solitons in the one-dimensional (1D)
version of the NLS equation with constant coefficients is a
well-established fact, which has important implications in various
areas of physics. In particular, solitons in fiber-optic
telecommunications \cite{Hasegawa} and quasi-1D Bose-Einstein
condensates (BECs) with attractive interactions between atoms
\cite{BECsoliton}, have drawn a great deal of interest.

A new class of dynamical problems, which also have a vast potential for
physical applications, emerges in the investigation of soliton dynamics in
extended versions of the NLS equation, in which coefficients are periodic
functions of the evolutional variable. A well-known example is a nonlinear
fiber-optic link subjected to \textit{dispersion management} (DM), which
implies that the dispersion coefficient periodically alternates between
positive and negative values. The DM links support a family of stable
temporal solitons (see, e.g., Refs. \cite{DM}, and also Ref. \cite{Ming}).
Somewhat similar is a \textit{waveguide-antiwaveguide} system, which can be
realized in the spatial domain (planar optical waveguides). In the latter
case, a light beam is transmitted through a periodic concatenation of
nonlinear waveguiding and antiwaveguiding segments \cite{Kaplan}.  A common
feature of the latter system with the DM is that the coefficient which
periodically jumps between positive and negative values also belongs to the
linear part of the equation.

Another technique that may be useful for optical telecommunications is
\textit{nonlinearity management} (NLM), which assumes that the coefficient
in front of the nonlinear term periodically changes its sign. An advantage
offered by the NLM is a possibility to compensate the nonlinear phase shift
accumulating in pulses due to the Kerr nonlinearity of the optical fiber
\cite{Pare'}. In practical terms, the NLM can be implemented by dint of
elements with a strong quadratic ($\chi ^{(2)}$) nonlinearity, which are
periodically inserted into the fiber link. The $\chi ^{(2)}$ elements can
emulate a negative Kerr effect through the cascading mechanism \cite{Uri}.
Various other schemes of NLM in fiber-optic links were considered, including
its combination with the DM, amplifiers, etc. \cite{NMfiber-general}. A
related scheme makes use of the NLM in soliton-generating lasers based on
fiber rings \cite{NMfiber-laser}. The NLM for spatial solitons, which
assumes alternation of self-focusing and self-defocusing nonlinear layers in
planar \cite{Javid} or bulk \cite{Isaac} waveguides, was introduced too.

All these systems may be regarded as examples of periodically
inhomogeneous optical waveguiding media. Other examples belonging
to the same general class are \textit{tandem waveguides} (see Ref.
\cite{tandem} and references therein) and the \textit{split-step}
\textit{model} \cite{SSM}. These are built as a juxtaposition of
linear segments alternating with ones featuring, respectively,
quadratic or cubic nonlinearity. A common feature of the models of
all these types is that they support \emph{robust} solitons,
despite a ``naive" expectation that solitons would quickly decay,
periodically hitting interfaces between strongly different
elements of which the system is composed.

The above-mentioned optical media are described by the NLS equation, in
which the role of the evolutional variable belongs to the propagation
distance, while the remaining free variable is either the local time (for
temporal solitons), or the transverse coordinate(s), in the spatial-domain
models. Mathematically similar, but physically altogether different, models
describe BECs in the 1D geometry. In that case, the corresponding NLS
equation is usually called the Gross-Pitaevskii (GP) equation. It governs
the evolution of the mean-field wave function $\phi $ in time ($t$), the
other variable, $x$, being the coordinate along the quasi-1D trap. In the
normalized form, the GP equation is
\begin{equation}
i\phi _{t}=-\frac{1}{2}\phi _{xx}+U(x)\phi +g|\phi |^{2}\phi ,  \label{GPE}
\end{equation}where $U(x)$ is the potential which confines the condensate, and the
nonlinearity coefficient $g$ is proportional to the scattering
length of collisions between atoms. Two natural possibilities to
introduce a time-periodic (ac) ``management" in the BEC context
are either through a periodic modulation of the confining
potential, most typically in the form of $U(x,t)=\frac{1}{2}\left[
\kappa _{0}+\kappa _{1}\cos (\omega t)\right] x^{2}$, or by means
of time modulation of the scattering length, using the Feshbach
resonance (FR) \cite{FR}. In the latter case, the nonlinearity
coefficient in Eq. (\ref{GPE}) takes the form of
$g(t)=g_{0}+g_{1}\sin (\omega t)$. In either case, the modulation
is generated by a combination of dc and ac magnetic fields applied
to the BEC.

The GP equation with the periodically modulated strength of the
trapping potential was considered for both $g>0$ (when solitons do
not exist, and the BEC as a whole is subjected to the
``management", including the 2D and 3D cases) \cite{GarciaGarcia},
and $g<0$, when the soliton is the basic dynamical object
\cite{Tashkent}. In particular, a parametric resonance is possible
in the former case, and creation of an effectively trapping
potential, while the underlying one is anti-trapping, having
$\kappa _{0}<0$, by the high-frequency ac part of the potential
(with large $\omega $) was predicted in the latter case.

The periodic modulation of the nonlinearity coefficient through
the ``ac FR management" is an especially interesting possibility,
as the FR is a highly efficient experimental tool, broadly used
for the study of various dynamical properties of the BECs
\cite{FR}. In particular, it has been predicted that the
modulation through the ac FR makes it possible to preclude
collapse and generate stable soliton-like structures in 2D (but
not 3D) condensates \cite{stabilization}; in fact, this prediction
is similar to the earlier considered possibility of the
stabilization of 2D spatial optical solitons in a bulk waveguide
subjected to the periodic NLM \cite{Isaac}. In the 1D model of the
GP type, subjected to the NLM, various stable dynamical states,
including Gaussian-shaped soliton-like objects, and ones of the
Thomas-Fermi type, were studied in detail \cite{Athens}. In
addition, analysis based on averaged equations was developed, for
this case, in Ref. \cite{DimDim} (similarly to the analysis
elaborated in Ref. \cite{Tashkent} for the case of the periodic
modulation of the trapping potential).

The objective of this work is to study \emph{resonance effects}
produced by the ac FR management, i.e., harmonic modulation of the
nonlinearity coefficient, in the dynamics of fundamental and
higher-order 1D solitons in the NLS equation. We will focus on the
case when the ac part of the nonlinear coefficient is small in
comparison with its constant (dc) part $g_{0}$, which accounts for
the self-attraction in the BEC, and is normalized to be
$g_{0}=-1$. We also assume that the soliton's width is much
smaller than the effective size of the trap, hence the external
potential may be dropped. The consideration of the GP equation
without the trapping potential makes it possible to identify
fundamental dynamical effects for the solitons induced by the ac
FR management. In this connection, it is necessary to mention
that, in the 1D case, the trapping potential is not a crucial
factor, on the contrary to the 2D and 3D cases, where the external
potential plays a much more important role, in view of the
intrinsic instability of the multi-dimensional NLS solitons.

Thus, we will be dealing with the normalized NLS\ equation in the form [cf.
Eq. (\ref{GPE})]
\begin{equation}
i\phi _{t}+\frac{1}{2}\phi _{xx}+\left[ 1+b\sin \left( \omega
t\right) \right] |\phi |^{2}\phi =0,  \label{NLSE}
\end{equation}where the amplitude $b$ of the ac drive is small. Note that Eq. (\ref{NLSE})
conserves exactly two dynamical invariants: the norm, which is proportional
to the number of atoms in the BEC,
\begin{equation}
N=\int_{-\infty }^{+\infty }\left\vert \phi (x)\right\vert ^{2}dx,  \label{N}
\end{equation}and the momentum,
\begin{equation}
P=i\int_{-\infty }^{+\infty }(\phi \phi _{x}^{\ast }-\phi ^{\ast }\phi
_{x})dx.  \label{P}
\end{equation}

We will demonstrate that the weak ac perturbation in Eq. (\ref{NLSE}) can
generate strong effects, if the driving frequency $\omega $ is close to
specific resonant values. These effects include intrinsic vibrations of the
fundamental soliton and splitting of the higher-order ones. We will also
propose analytical explanations to these effects. To the best of our
knowledge, these results have not been reported before for the present
simple model.

The rest of the paper is organized in the following way. The resonant effect
of the periodic NLM on the fundamental soliton is reported in Section 2, and
the resonant splitting of $n$-solitons with $n=2$ and $3$ is investigated in
Section 3. Section 4 concludes the paper.

\section{Resonant response of the fundamental soliton}

\subsection{Numerical results}

First, we consider the action of the ac perturbation in Eq. (\ref{NLSE}) on
the fundamental soliton, which, in the case $b=0$, is
\begin{equation}
\phi _{\mathrm{sol}}(x,t)=A~\mathrm{sech}(A\left( x-x_{0}\right) )\exp
\left( iA^{2}t/2\right) ,  \label{sol}
\end{equation}where $A$ is an arbitrary amplitude. Numerical simulations of Eq. (\ref{NLSE}
) were performed in a sufficiently large domain, $0<x<L$, the
initial condition corresponding to the soliton (\ref{sol}) placed
at the center of the domain, $x_{0}=L/2$. We have performed
numerical simulations using the split-step Fourier method with
1024 Fourier modes. The system size is $L=50$, and the time step
for the numerical simulation is $\Delta t=0.001$.

\begin{figure}[tbph]
\begin{center}
\includegraphics[width=3.5in]{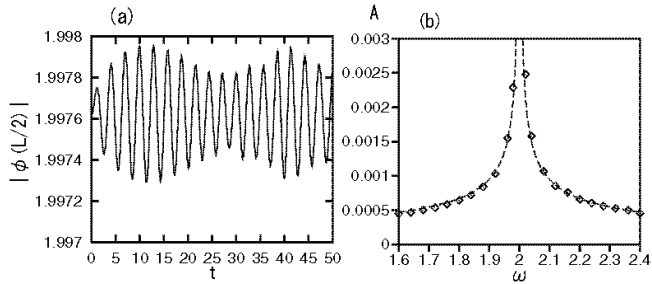}
\end{center}
\caption{(a) A typical example of beatings in the evolution of the
fundamental-soliton's amplitude under the action of a very weak
perturbation in Eq. (\protect\ref{NLSE}), with $b=10^{-4}$ and
$\protect\omega =2.2$. The amplitude of the initial unperturbed
soliton is $A=2$ (the corresponding soliton's frequency is
$\protect\omega _{\mathrm{sol}}\equiv A^{2}/2=2$). (b) The
difference between the maximum and minimum values of the soliton's
amplitude vs. the perturbation frequency. The dashed line is a
fitting curve, $0.0003\cdot |\protect\omega -2|^{-1/2}$.}
\label{fig1}
\end{figure}

Figure \ref{fig1}(a) displays a typical example of the time evolution of the
soliton's amplitude $|\phi (x=L/2)|$, under the action of the ac
perturbation with a very small amplitude, $b=0.0001$. The frequency of the
beatings observed in this figure can be clearly identified as $\omega
-\omega _{\mathrm{sol}}$, where $\omega _{\mathrm{sol}}\equiv A^{2}/2$ is
the intrinsic frequency of the unperturbed soliton (\ref{sol}). We have
checked that the the beating frequency is independent of the system's size
and the other details of the numerical scheme.

\begin{figure}[tbph]
\begin{center}
\includegraphics[width=2.2in]{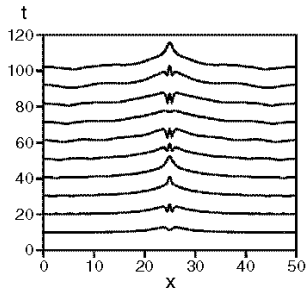}
\end{center}
\caption{
Direct numerical solution of the linearized equation
(\protect\ref{lin}) for the perturbation around the fundamental
soliton, in the near-resonance case, with $\protect\omega =1.98$.}
\label{fig2}
\end{figure}

The main resonant effect for the fundamental soliton is displayed
in Fig. \ref{fig1}(b), in the form of the difference between the
maximum and minimum (in time) values of its amplitude versus the
driving frequency $\omega $. The resonance at $\omega =\omega
_{\mathrm{sol}}=2$ (for $A=2$) is obvious. The simulations do not
reveal any noticeable subharmonic or higher-order resonance at
frequencies $\omega =1$, $3$ or $4$. Due to the scaling invariance
of Eq. (\ref{NLSE}), the plot shown in Fig. \ref{fig1}(b) does not
pertain solely to the particular value of the soliton's amplitude,
$A=2$, but is actually a universal one. It is easy to verify that
the ranges of the variables $t$ and $x$, which are shown in this
and other figures, correspond, in the normalized units, to
experimentally realistic configurations of the BECs in the
quasi-1D geometry.

\subsection{Perturbative analysis}

In order to explain the resonance shown above, we look for a
perturbed fundamental solution as $\phi (x,t)=\phi
_{\mathrm{sol}}(x,t)+\phi _{\mathrm{pert}}(x,t)$, where the first
term is the solution (\ref{sol}). Thus, we arrive at the driven
linearized equation for the perturbation,
\begin{eqnarray}
&&i\left( \phi _{\mathrm{pert}}\right) _{t}+\frac{1}{2}\left( \phi _{\mathrm{\ pert}}\right) _{xx}+A^{2}\mathrm{sech}^{2}(Ax)\left( 2\phi _{\mathrm{pert}}+e^{iA^{2}t}\phi _{\mathrm{pert}}^{\ast }\right)  \nonumber \\
&=&\frac{i}{2}bA^{3}\mathrm{\ \ sech}^{3}(Ax)\left[ e^{i\left( \left(
A^{2}/2\right) +\omega \right) t}-e^{i\left( \left( A^{2}/2\right) -\omega
\right) t}\right] .  \label{lin}
\end{eqnarray}The source of the resonant response is in the fact that the second term on
the right-hand side of Eq. (\ref{lin}) becomes time-independent
exactly at the resonance point, $\omega =\omega
_{\mathrm{sol}}\equiv A^{2}/2$. Figure \ref{fig2} displays the
evolution of the perturbation close to the resonance at $\omega
=1.98$, as found from direct numerical integration of the
linearized equation (\ref{lin}), with the initial condition $\phi
_{\mathrm{pert}}(x)=0$. As can be seen from the figure, the
perturbation grows in time at the center, and simultaneously
expands in space. Strictly speaking, the latter feature remains
valid as long as the size of the region occupied by the expanding
wave fields remains essentially smaller than the limit imposed by
the confining field.

The linearized equation (\ref{lin}) is too difficult for an exact
analytical solution. However, the observation that the
characteristic spatial scale of the solution observed in Fig.
\ref{fig2} becomes much larger than the internal scale of the
function $\mathrm{sech}(Ax)$ suggests that principal features of
the solution can be understood from a simpler equation, in which
the term $\sim \mathrm{sech}^{2}(Ax)$ on the left-hand side of Eq.
(\ref{lin}) is neglected, and the source corresponding to the
second term on the right-hand side is approximated by a $\delta
$-function:
\begin{equation}
i\left( \tilde{\phi}_{\mathrm{pert}}\right) _{t}+\frac{1}{2}\left(
\tilde{\phi}_{\mathrm{pert}}\right) _{xx}=\mathrm{const}\cdot
\delta (x)e^{-i\Delta \omega \cdot t},~\mathrm{const}=\frac{i\pi
}{4}bA^{2},  \label{simple}
\end{equation}where $\Delta \omega \equiv \omega -\omega _{\mathrm{sol}}$, and $\mathrm{const}\equiv \left( ib/2\right) A^{3}\int_{-\infty }^{+\infty }\mathrm{sech}^{3}(Ax)dx$. Equation (\ref{simple}) can be solved by means of the Fourier
transform. After straightforward manipulations, this
yields\begin{equation}
\tilde{\phi}_{\mathrm{pert}}(x,t)=-\frac{\mathrm{const}}{2\sqrt{\pi
}}(1+i)e^{-i\Delta \omega \cdot t}\int_{0}^{t}\frac{dt^{\prime
}}{\sqrt{t^{\prime }}}\exp \left( \frac{ix^{2}}{2t^{\prime
}}+i\Delta \omega \cdot t^{\prime }\right) .  \label{solution}
\end{equation}Further consideration shows that, for $\Delta \omega <0$, the asymptotic
form of the solution (\ref{solution}) at $t\rightarrow \infty $ amounts to
an exponentially localized stationary expression, which, by itself, is an
exact solution to Eq. (\ref{simple}):
\begin{equation}
\tilde{\phi}_{\mathrm{pert}}(x,t)=\frac{\mathrm{const}}{\sqrt{2\left\vert
\Delta \omega \right\vert }}\exp \left( i\left\vert \Delta \omega
\right\vert t-\sqrt{2\left\vert \Delta \omega \right\vert }|x|\right) .
\label{stationary}
\end{equation}In the case of $\Delta \omega >0$, the asymptotic form of the general
solution (\ref{solution}) corresponds to a symmetric region
occupied by plane waves emitted by the central source at
wavenumbers $k=\pm \sqrt{2\Delta \omega }$. The region expands in
time with the group velocities $v_{\mathrm{gr}}=k=\pm
\sqrt{2\Delta \omega }$, so that the asymptotic form of the
solution is
\begin{equation}
\tilde{\phi}_{\mathrm{pert}}(x,t)\approx
-i\frac{\mathrm{const}}{\sqrt{2\Delta \omega }}\cdot \left\{
\begin{array}{cc}
\exp \left( -i\Delta \omega \cdot t+i\sqrt{2\Delta \omega }|x|\right) , &
\mathrm{if}~|x|~<\sqrt{2\Delta \omega }t, \\
0, & \mathrm{if}~|x|~>\sqrt{2\Delta \omega }t.\end{array}\right.
\label{expanding}
\end{equation}This asymptotic solution implies that the norm (\ref{N}) of the expanding
radiation field grows in time at the rate
$dN_{\mathrm{pert}}/dt=\sqrt{2/\Delta \omega }\left\vert
\mathrm{const}\right\vert ^{2}$, which, in fact, is the rate at
which the norm flows from the soliton to the radiation waves
emitted under the action of the ac perturbation.

Both analytical expressions (\ref{stationary}) and
(\ref{expanding}) feature the $\left\vert \Delta \omega
\right\vert ^{-1/2}$ factor, that perfectly fits the numerical
data summarized in Fig. \ref{fig1}(b). Although these results,
obtained for the weak time-periodic FR management, seem very
simple, they have not been reported before, to the best of our
knowledge. We also notice that the usual variational approximation
(VA)\ for the NLS solitons, which is efficient in explaining a
number of other perturbative effects \cite{VA}, cannot account for
the occurrence of the resonance at $\omega =\omega
_{\mathrm{sol}}$, because the VA neglects radiation effects, while
the above consideration showed that it is exactly the radiation
field which is amenable for the manifestations of the resonance.

The above results were obtained in the linear approximation, i.e.,
for a very small amplitude $b$ of the ac drive in Eq.
(\ref{NLSE}). At larger $b$, the perturbed soliton can either
survive or decay into radiation. In fact, a stability region for
the solitons in a similar model with a nonsmall perturbation,
which differs from that in Eq. (\ref{NLSE}) by the form of the
periodic modulation function, which is a piecewise-constant one,
rather than harmonic, was drawn in Ref. \cite{Javid} in the
context of a model for spatial optical solitons in a layered
waveguide. In the cases when a stable soliton established itself
in the strongly perturbed (``strongly nonlinearly-managed")
system, its formation from the initial configuration (\ref{sol})
went through emission of radiation and, sometimes, separation of a
small secondary pulse, while no pronounced resonance at $\omega
=\omega _{\mathrm{sol}}$ was observed. As we do not expect that
the replacement of the piecewise-constant modulation function by
the harmonic one should dramatically alter the stability region,
we do not consider this issue here in detail.

\section{Resonant splitting of higher-order solitons}

\subsection{Response to a very weak ac drive}

As is well known, the unperturbed NLS equation gives rise to exact soliton
solutions of order $n$, in the form of periodically oscillating breathers,
which start from the initial conditions
\begin{equation}
\phi _{0}(x)=nA~\mathrm{sech}\left( A\left( x-L/2\right) \right)  \label{n}
\end{equation}with an integer $n>1$ \cite{SatsumaYajima} [the expression (\ref{n}) assumes
that the initial configuration is placed at the center of the integration
domain]. The frequency of the shape oscillations (breathings) of the
higher-order soliton is
\begin{equation}
\omega _{\mathrm{br}}=4A^{2},  \label{omega_br}
\end{equation}irrespective of the value of $n$.

Generally speaking, the higher-order solitons are unstable bound
complexes of fundamental solitons, as, in the absence of
perturbations, their binding energy is exactly zero, which is a
known consequence of the exact integrability of the unperturbed
NLS equation. Nevertheless, not any perturbation readily splits
the higher-order soliton into its fundamental constituents;
usually, the splitting is easily induced by specific
nonconservative terms added to the NLS equation, such as the one
accounting for the intra-pulse stimulated Raman scattering in
optical fibers \cite{Hasegawa}. The consideration of dynamics of
the higher-order solitons is also relevant, especially in the
context of BECs, as the corresponding initial configurations can
be created in the real experiment.

\begin{figure}[tbph]
\begin{center}
\includegraphics[width=5in]{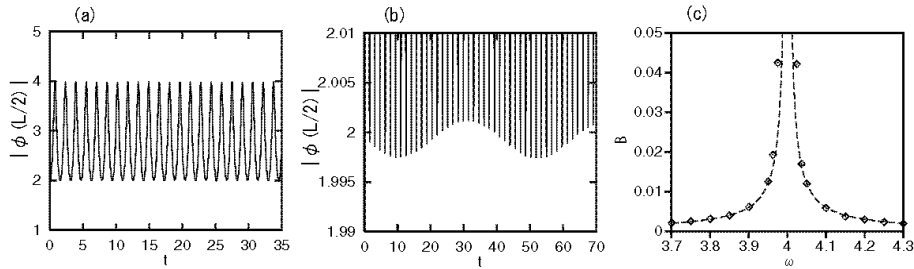}
\end{center}
\caption{(a) Oscillations of the amplitude, $\left\vert
\protect\phi (x=L/2)\right\vert $, of the $2$-soliton created by
the initial condition (\protect\ref{n}) with $A=1$, in the case
of $b=0.5\times 10^{-4}$ and $\protect\omega =4.15$. (b) Zoom of
the previous panel around minimum values of the amplitude, which
reveals beatings at the frequency $\protect\omega -\protect\omega
_{\mathrm{br}}$. (c) The difference between the maximum and
minimum values of the soliton's amplitude vs. the driving
frequency $\protect\omega $. The dashed line is a fitting curve,
$0.0006/|\protect\omega -4|$; note the difference of the fitting
power, $-1$, from that, $-1/2$, in Fig. \protect\ref{fig1}(b).}
\label{fig3}
\end{figure}

We have studied in detail the $n$-solitons up to $n=5$. First, we
consider the case of a very small driving amplitude, $b=0.00005$.
Figure \ref{fig3} (a) displays oscillations of the amplitude
$|\phi (x=L/2)|$ of the $2$ -soliton, which corresponds to the
initial condition (\ref{n}) with $n=2$. The frequency of the basic
oscillations coincides with $\omega _{\mathrm{br}} $, as given by
the expression (\ref{omega_br}), while the frequency of the zoomed
beatings in Fig. \ref{fig3}(b) can be clearly identified with
$\omega -\omega _{\mathrm{br}}$. The resonant character of the
response of the $2$-soliton to the weak NLM is obvious from Fig.
\ref{fig3}(c).

Note that the exact solution for the $n$-soliton features not only
the shape-oscillation frequency (\ref{omega_br}), but also an
overall frequency of the phase oscillations, which coincides with
the above-mentioned frequency $\omega _{\mathrm{sol}}=A^{2}/2$ for
the fundamental soliton, provided that the initial condition is
taken as in Eq. (\ref{n}). In the simulations, we also observed a
resonant response at $\omega =\omega _{\mathrm{sol}}$, but this
resonance was essentially weaker than the one at $\omega =\omega
_{\mathrm{br}}$. In particular, this is manifest in the fact that,
as well as in the case of the fundamental soliton, the fit to the
response around the former resonance is provided by the expression
$\left\vert \omega -\omega _{\mathrm{sol}}\right\vert ^{-1/2}$,
cf. Fig. \ref{fig1}(b), while the fit to the resonance at $\omega
=\omega _{\mathrm{sol}}$ demonstrates a more singular dependence,
$\sim \left\vert \omega -\omega _{\mathrm{br}}\right\vert ^{-1}$,
as seen in Fig. \ref{fig3}(c). Another qualitative difference
between the two resonances is that the one at $\omega =\omega
_{\mathrm{br}}$, with a larger (but still small) forcing parameter
$b $, leads to splitting of the higher-order solitons into
fundamental ones, as shown below, while, in the case of the
resonance at $\omega =\omega _{\mathrm{sol}}$, the increase of $b$
does not lead to the splitting.

\subsection{Splitting of $2$- and $3$-solitons}

Unlike the case of the fundamental soliton, the reaction of the
higher-order ones to larger values of the forcing parameter was
not studied before, therefore we have done it here. First, we aim
to demonstrate that the $2$-soliton readily splits into two moving
fundamental pulses, if the driving frequency is close to the
resonant value (\ref{omega_br}). The shape of each moving soliton
is very close to that given by the commonly known exact solution,
which can be obtained by application of the Galilean boost, with a
velocity $v$, to the zero-velocity fundamental soliton
(\ref{sol}),
\begin{equation}
\phi _{\mathrm{sol}}(x,t)=A~\mathrm{sech}(A(x-vt))\exp \left[ ivx+\left(
i/2\right) \left( A^{2}-v^{2}\right) t\right] .  \label{boost}
\end{equation}

\begin{figure}[tbph]
\begin{center}
\includegraphics[width=3.5in]{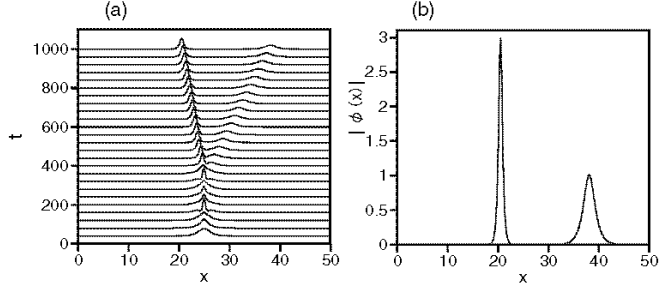}
\end{center}
\caption{A typical example of the splitting of a $2$-soliton [generated by
the initial condition (\protect\ref{n}) with $n=2$ and $A=1$] into an
asymmetric pair of moving fundamental solitons, under the action of the weak
resonant drive, with $\protect\omega =4$ and $b=0.0005$. (a) The evolution
of $\left\vert u(x,t)\right\vert $. (b) The final configuration at $t=1000$.}
\label{fig4}
\end{figure}

Figure \ref{fig4} displays the evolution of the wave function for
the initial condition (\ref{n}) with $n=2$ in the resonant case
($A=1$ and $\omega =4$), with the driving amplitude $b=0.0005$.
The latter value is still very small, but larger by a factor of
$10$ than in the case shown in Fig. \ref{fig3}. The amplitudes of
the two fundamental solitons, observed as a result of the
splitting, are close to $A_{1}=3$ and $A_{2}=1$ [note that they
exactly corresponds to the fundamental-soliton constituents of the
original $2$-soliton with $A=1$, in terms of the inverse
scattering transform (IST) \cite{SatsumaYajima}]. Velocities of
the splinters were measured to be $v_{1}=0.00197$ and
$v_{2}=0.0066$, respectively (with the ratio $v_{1}:v_{3}\approx
1:3$). At the end of the simulation run ($t=1000$ ), the secondary
solitons are found at the distance, respectively, $4.5$ and $13.2$
from the central point, $x=L/2$.

\begin{figure}[tbph]
\begin{center}
\includegraphics[width=3.5in]{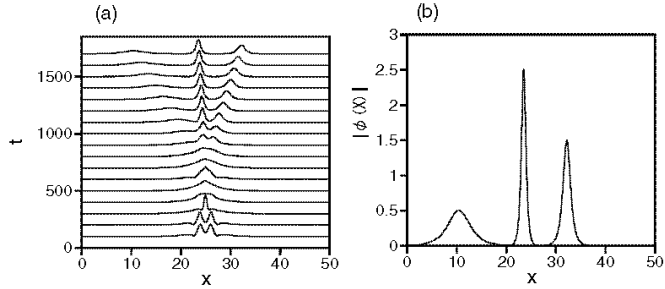}
\end{center}
\caption{The same as in Fig. \protect\ref{fig4} for the
$3$-soliton, generated by the initial configuration
(\protect\ref{n}) with $n=3$ and $A=0.5$. In this case, the
forcing frequency and amplitude are $\protect\omega =1$ and
$b=0.0005$.} \label{fig5}
\end{figure}

Similar near-resonant splittings were observed for $n$-solitons
with $n>2$. In particular, Fig. \ref{fig5} shows this outcome for
$n=3$, which corresponds to the initial configuration (\ref{n})
with $n=3$, $A=0.5$, $\omega =1$ and $b=0.0005$. This time, the
splitting gives rise to three moving fundamental solitons, whose
amplitudes are close to $A_{1}=2.5$, $A_{2}=1.5$, and $A_{3}=0.5$.
As well as in the case of $n=2$, these values correspond to the
constituents of the original $3$-soliton (with $A=0.5$), in terms
of the IST \cite{SatsumaYajima}. The velocities of the three
splinters are $v_{1}=-0.00146$, $v_{2}=0.0732$, and
$v_{3}=-0.0148$, so that the ratios between them are
$v_{1}/v_{2}\approx -1/5$ and $v_{3}/v_{2}\approx -2$.

\begin{figure}[tbph]
\begin{center}
\includegraphics[width=3.5in]{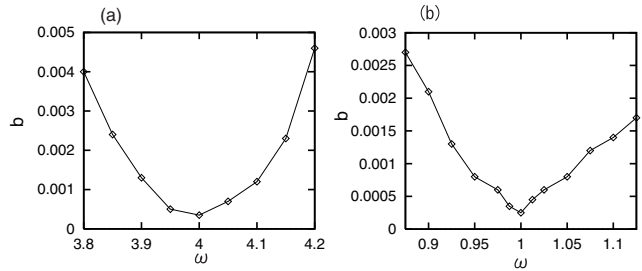}
\end{center}
\caption{The minimum values of the amplitude of the ac perturbation,
necessary for the splitting of the $2$-soliton (a) and $3$-soliton (b), as
functions of the driving frequency. The initial condition is taken in the
form of Eq. (\protect\ref{n}) with, respectively, $n=2$ and $A=1$, or $n=3$
and $A=0.5$. In both cases, the sharp minimum exactly corresponds to the
resonant frequency, as predicted by Eq. (\protect\ref{omega_br}).}
\label{fig6}
\end{figure}

These results can be summarized in the form of diagrams which show
the minimum (threshold) value of the forcing amplitude $b$,
necessary for the splitting, versus the driving frequency $\omega
$. The splitting of the $2$- and $3$-solitons was registered if it
took place in the simulations of Eq. (\ref{NLSE}) that were run up
to the time, respectively, $t=600$ or $t=2000$ (still longer
simulations did not give rise to any essential difference in the
results). As is seen from Fig. \ref{fig6}, for both $2$- and
$3$-solitons these dependences clearly have a resonant shape, with
sharp minima at the frequency given by Eq. (\ref{omega_br}). It is
not quite clear why the forcing amplitude required for the
splitting is very small but finite even exactly at the resonance
point. This may be related to the accuracy of the numerical scheme
and/or the finite size of the integration domain. Similar
observations were also made in simulations of the $n$-solitons
with $n=4$ and $5$.

\subsection{Analytical results}

The amplitudes and velocities of the fundamental solitons, into which the
higher-order ones split, can be predicted in an analytical from. As it was
already mentioned above, the amplitudes of the secondary solitons coincide
with those which correspond to the constituents (eigenvalues) of the
corresponding original $n$-soliton in terms of the IST. However, the
velocities of the emerging fundamental solitons cannot be forecast this way,
as, in terms of the IST, they are zero when the fundamental solitons are
bound into a higher-order one.

Nevertheless, both the amplitudes and velocities of the final set of the
solitons can be predicted in a different way, using the exact and nearly
exact conservation laws of Eq. (\ref{NLSE}). Indeed, there are two exact
dynamical invariants, (\ref{N}) and (\ref{P}), and, in addition to that, the
unperturbed NLS equation has an infinite series of higher-order dynamical
invariants, starting from the Hamiltonian,
\begin{equation}
H=\frac{1}{2}\int_{-\infty }^{+\infty }(|\phi _{x}|^{2}-|\phi |^{4})dx.
\label{H}
\end{equation}Two next invariants, which do not have a straightforward physical
interpretation, are \cite{Zakharov}
\[
I_{4}=\frac{1}{2}\int_{-\infty }^{+\infty }\left( \phi \phi _{xxx}^{\ast
}+3\phi \phi _{x}^{\ast }|\phi |^{2}\right) dx,
\]\begin{equation}
I_{5}=\frac{1}{4}\int_{-\infty }^{+\infty }\left[ |\phi _{xx}|^{2}+2|\phi
|^{6}-\left( \left( |\phi |^{2}\right) _{x}\right) ^{2}-6|\phi
_{x}|^{2}|\phi |^{2}\right] dx.  \label{I5}
\end{equation}

In the case of the splitting of the $2$-soliton (\ref{n}) with the
amplitude $A$, the exact conservation of the norm (\ref{N}) and
approximate conservation of the Hamiltonian (\ref{H}) yield the
following relations between $A$ and the amplitudes $A_{1,2}$ of
the emerging fundamental solitons (splinters): $4A=A_{1}+A_{2}$,
and $28A^{3}\approx A_{1}^{3}+A_{2}^{3}$ (the latter relation
neglects small kinetic energy of the emerging solitons). These two
relations immediately yield $A_{1}=3A$ and $A_{2}=A$, which
coincides with the the above-mentioned numerical results, as well
as with the predictions based on the set of the $2$-soliton's IST\
eigenvalues. Furthermore, the exact momentum conservation yields a
relation involving the velocities $v_{1,2}$ of the secondary
solitons, $A_{1}v_{1}+A_{2}v_{2}=0$. With regard to the ratio
$A_{1}/A_{2}=3$, this implies $v_{1}/v_{2}=-A_{2}/A_{1}=-1/3$.
This relation is indeed consistent with the aforementioned
numerical results, although the absolute values of the velocities
cannot be predicted this way.

Similarly, in the case of the splitting of the $3$-soliton, the
exact conservation of $N$ and approximate conservation of $H$ and
$I_{5}$ [see Eq. (\ref{I5})] yield the relations (which again
neglect small kinetic terms, in view of the smallness of the
observed velocities) $9A=A_{1}+A_{2}+A_{3}$, $153A^{3}\approx
A_{1}^{3}+A_{2}^{3}+A_{3}^{3}$, and $3369A^{5}\approx
A_{1}^{5}+A_{2}^{5}+A_{3}^{5}$. A solution to this system of
algebraic equations is $A_{1}=5A$, $A_{2}=3A$, $A_{3}=A$, which
are the same values that were found from the direct simulations,
and can be predicted as the IST eigenvalues. The conservation of
$P$ and $I_{4}$ gives rise to further relations,
$A_{1}v_{1}+A_{2}v_{2}+A_{3}v_{3}=0$ and
$(A_{1}v_{1}^{3}-A_{1}^{3}v_{1})+(A_{2}v_{2}^{3}-A_{2}^{3}v_{2})+(A_{3}v_{3}^{3}-A_{3}^{3}v_{3})=0
$. If the velocities $v_{1,2}$ are small, it follows from here
that
$v_{1}/v_{2}=-(A_{2}^{3}-A_{2}A_{3}^{2})/(A_{1}^{3}-A_{1}A_{3}^{2})=-1/5$,
and
$v_{3}/v_{2}=-(A_{2}^{3}-A_{2}A_{1}^{2})/(A_{3}^{3}-A_{3}A_{1}^{2})=-2$.
These results for the velocities are consistent with the numerical
situation observed in Fig.~5.

\section{Conclusion}

In this work, we have addressed a simple model, based on the NLS equation,
which describes an attractive Bose-Einstein condensate (BEC) in a quasi-1D
trap, with the nonlinearity strength subjected to a weak time-periodic (ac)
modulation (that can be imposed by means of the Feshbach-resonance
technique). The same model describes the nonlinearity management in
periodically inhomogeneous optical waveguides.

It was found from direct simulations, and explained by means of a
straightforward perturbative expansion, that the response of a
fundamental soliton, in the form of temporal beatings of its
amplitude, to the weak ac perturbation is resonant when the
driving frequency $\omega $ is close to the soliton's intrinsic
frequency. For $n$-solitons (breathers), with $n=2$ and $3$, the
response to an extremely weak drive is also resonant, if $\omega $
is close to the breathing frequency. More interestingly, a
slightly stronger drive gives rise to splitting of the $2$- and
$3$-solitons into sets of two or three moving fundamental
solitons, respectively. The dependence of the minimum perturbation
amplitude, which is necessary for the splitting, on $\omega $ has
a clearly resonant character too. The amplitudes of the splinter
solitons, and the ratio of their velocities, can be easily
predicted on the basis of the exact and approximate conservation
laws of the perturbed NLS equation.

\section*{Acknowledgment}

The work of B.A.M. was partially supported by the grant No. 8006/03 from the
Israel Science Foundation.

\end{document}